\documentstyle[fleqn,twoside, espcrc2]{article}

\input{tcilatex}
\begin{document}

\title{{\tt www.fermiqcd.net}}
\author{Massimo~Di~Pierro\address[depaul]{
School of Computer Science, Telecommunications and Information 
Systems, DePaul University, Chicago, IL 60604}\thanks{presented by
M.~Di~Pierro},
Aida~X.~El-Khadra\address[urbana]{
Department of Physics, University of Illinois, Urbana, IL 61801},
Steven Gottlieb\address[indiana]{
Department of Physics, Indiana University, Bloomington, IN 47405}
Andreas~S.~Kronfeld\address[fermilab]{
Fermi National Accelerator Laboratory, P.O.~Box~500, 
Batavia, IL 60510},                                  
Paul~B.~Mackenzie\addressmark[fermilab], 
Masataka~Okamoto\addressmark[fermilab],
Mehmet~B.~Oktay\addressmark[urbana] and 
James~N.~Simone\addressmark[fermilab] 
(for the {\bf FermiQCD Collaboration})}
\maketitle

\section{INTRODUCTION}

FermiQCD\cite{mdp}\cite{fermiqcd} is a C++ library for fast development of
parallel lattice QCD applications~\footnote{%
Ref. \cite{flynn}, for example ,is entirely based on FermiQCD.}. 
The expression FermiQCD Collaboration is
used as a collective name to indicate both the users of the software and
its contributors. 

One of the main differences between FermiQCD and libraries developed by
other collaborations is that it follows an object oriented design as
opposed to a procedural design. FermiQCD should not be identified
exclusively with the 
implementation of the algorithms but, rather, with the strict
specifications that define its Application Program Interface. One
should think of FermiQCD as a language on its own (a superset of the C++
language), designed to describe Lattice QCD algorithms. The objects of the
language include complex numbers (mdp\_complex), matrices (mdp\_matrix),
lattices (mdp\_lattice), fields (gauge\_field, fermi\_field,
staggered\_field), propagators (fermi\_propagator) and actions. Algorithms
written in terms of these objects are automatically parallel.

Some of the advantages of our design approach are the following:

\begin{itemize}
\item  Programs written in FermiQCD are easy to write, read and modify since
the FermiQCD syntax resembles the mathematical syntax used in Quantum Field
Theory articles and books.

\item  Programs are portable in the sense that they can, in principle, be
compiled with any ANSI C++ compiler. Hardware specific optimizations are
coded in the library and hidden from the high level programmer.

\item  The high level programmer does not have to deal with
parallelization issues since the underlying objects deal with it.
FermiQCD communications are based on MPI. 

\item  Programs are easier to debug because the usage of FermiQCD 
objects and algorithms does not require explicit use of pointers. All 
memory management
is done by the objects themselves.
\end{itemize}

FermiQCD was originally designed with one
main goal in mind: easy of use. It is true that in many cases this 
requires a compromise with speed. For example, FermiQCD actions and
fields support arbitrary $SU(n_c)$ gauge group and it is not practical to
optimize the linear algebra for any $n_c$. However it is convenient
to optimize some of the specific cases of interest (such as $n_c=3$) while
maintaining compatibility with the FermiQCD syntax. At present the
FermiQCD library includes multiple implementations of each action (Wilson,
Clover, Kogut-Susskind, Asqtad\cite{asqtad}, Domain Wall, Fermilab\cite
{fermilab}) and, for each action, at least one implementation is optimized
for $SU(3)$.
In particular, we provide a FermiQCD port of the Clover $SU(3)$ action
coded by M.~L\"{u}scher using SSE assembly instructions 
for Pentium 4\cite{luscher}.

The following code declares two fermionic fields ({\tt phi} and 
{\tt psi}) and one $SU({\tt nc})$ gauge field ({\tt U}) on a given lattice (%
{\tt lattice}), reads {\tt psi} and {\tt U} from disk and computes 
\[
\varphi =Q[U,\kappa ]^{-1}\psi 
\]
where $Q[U,\kappa ]=D[U]+m[\kappa ]$ is the Dirac matrix and $\kappa $ is
the typical lattice parameter that sets the mass scale:
\begin{verbatim}
fermi_field phi(lattice, nc);
fermi_field psi(lattice, nc);
gauge_field U(lattice,nc);
coefficients light_quark;
psi.load("fermi_field.mdp");
U.load("gauge_field.mdp");
light_quark["kappa"]=0.123;
if(nc==3) default_fermi_action=
          FermiCloverActionSSE2::mul_Q;
mul_invQ(phi,psi,U,light_quark);
\end{verbatim}\vskip -2mm
Note that when ${\tt nc}=3$ the above program uses the SSE optimized Wilson
action. If ${\tt nc}\neq 3$ a different implementation
of the Wilson action
is used. FermiQCD programs are transparent to the choice of the action.

On a cluster of 2GHz Pentium 4 PCs connected by Myrinet, one double
precision SU(3) minimum residue step with Clover action takes $4.8$ micro
seconds per lattice site. The efficiency drops to $75-80$\% on 8
processors. For more benchmarks we refer to the {\tt www.fermiqcd.net} web
page. 

FermiQCD programs are implicitly parallel. Each lattice object 
determines an optimal communication pattern for its sites assuming a 
next-neighbor, a next-next-neighbor or a next$^3$-neighbor 
interaction in the action. The optimal patterns are determined according 
to empirical rules that minimize dependence on the network latency 
and minimize communication load.

\section{EXAMPLES}

\subsection{Computing the average plaquette}

Given a $\mu \nu $-plane, the average plaquette, for each gauge
configuration, is defined as
\begin{eqnarray}
\left\langle P_{\mu \nu }\right\rangle=\frac 1V Re Tr &\!\!\!\sum_x 
&\!\!\! U_\mu
(x)U_\nu (x+\widehat{\mu }) \\ 
&&\!\!\! \left[ U_\nu (x)U_\mu (x+\widehat{\nu })\right] ^H
\label{P}
\end{eqnarray}
here is the corresponding FermiQCD syntax:
\begin{verbatim}
forallsites(x)
  p=p+real(trace(U(x,mu)*U(x+mu,nu)*
       hermitian(U(x,nu)*U(x+nu,mu))));
p=p/lattice.size();
\end{verbatim}\vskip -2mm
In the above code {\tt forallsites} is a parallel loop, {U(x,mu)} is 
$n_c\times n_c$ color matrix.

\subsection{Implementing the Dirac-Wilson action}

As another example, we consider here the following Wilson discretization of
the Dirac action:
\begin{eqnarray}
\varphi _\alpha \!\!=\!\!\psi _\alpha \!\!-\!\!\kappa &\!\!\!\sum_\mu 
&\!\!\!\!\!(1\!+\!\gamma 
_\mu 
)_{\alpha
\beta }U_\mu (x)\psi _\beta (x\!+\!\widehat{\mu })+ \\
&&\!\!\!\!\!(1\!-\!\gamma _\mu )_{\alpha\beta }U_\mu 
^H(x\!-\!\widehat{\mu })\psi 
_\beta 
(x\!-\!\widehat{\mu })
\label{Action}
\end{eqnarray}
which can be translated into the following FermiQCD code:
\begin{verbatim}
phi=psi;
forallsites(x)
  for(int mu=0; mu<4; mu++) {
    for(int b=0; b<4; b++) {
      psi_up(b)=U(x,mu)*psi(x+mu,b);
      psi_dw(b)=hermitian(U(x-mu,mu))*
                   psi(x-mu,b);
    }
    phi=phi-kappa*((1+Gamma[mu])*psi_up+
                   (1-Gamma[mu])*psi_dw);
  }
\end{verbatim}\vskip -2mm
Note how ``{\tt 1}" in this context is interpreted as a diagonal unitary 
matrix. The sum on $\alpha$ is implicit since {\tt psi\_up} and {\tt 
psi\_dw} are spin$\times$color matrices. 

\subsection{Computing the pion propagator}

The pion propagator is defined as 
\begin{eqnarray}
C_2(x_0) &=&\sum_{x_1,x_2,x_3}\left\langle \pi (x)\pi (0)\right\rangle  \\
&=& Re Tr\sum_{x_1,x_2,x_3}\sum_{\alpha ,\beta }S_{\alpha \beta
}(x)S_{\beta \alpha }^H(x)  \label{prop}
\end{eqnarray}
where $S$ is a light quark propagator in the background gauge field U: 
\begin{equation}
S_{\alpha \beta }^{ij}(x)=\int [dq d\bar q]
q(x)\overline{q}(0)e^{Action[q,U]}
\label{S}
\end{equation}
Here is the FermiQCD syntax for Eq.~(\ref{prop}):
\begin{verbatim}
generate(S,U,light_quark);
forallsites(x) 
  for(int a=0; a<4; a++)
    for(int b=0; b<4; b++)
      C2(x(0))+=real(trace(S(x,a,b)*
                 hermitian(S(x,b,a))));
\end{verbatim}\vskip -2mm
where {\tt S(x,a,b)} is a color$\times$color matrix. The function 
{\tt generate} creates the propagator {\tt S} by calling the inverter
(for example the minimum residue or the stabilized bi-conjugate gradient)
. Each of the inverters works on any of the provided actions and also on
user defined actions.

\section{CONCLUSIONS}

In our experience the cost of developing and debugging software is one of
the major costs of Lattice QCD computations. 
FermiQCD was designed to
standardize this software development process and, therefore, reduce the
cost for the community. 

FermiQCD comes in the form of a library (with
examples), rather than a collection of ready made programs, since we
understand that different research groups have different requirements. 
Some ready
made programs for specific computations (such as generating gauge
configurations, computing meson propagators, etc.) are available as
examples, while others are available upon request to the authors.
We provide converters for the most common data formats including
UKQCD, MILC, CANOPY and various binary formats.
 
The FermiQCD libraries can be used
freely for research purposes and we do not require that users share their
programs unless they wish to do so (although we feel that this practice
should be encouraged). 

On one hand, FermiQCD is a mature project. The present implementation
has been tested independently by different groups and it has been used to
develop large scale lattice QCD computations. On the other hand, there is
a lot of work that needs to be done, including:

\begin{itemize}
\item  implementing dynamical fermions

\item  developing a graphical interface

\item  optimizing the gauge actions

\item  porting the low level communications libraries (currently based on
MPI) to other protocols such as TCP/IP for Gigabit ethernet and SciDAC API
for the QCDOC.

\item  incorporating grid-like features, including the ability to read and
write data in the new international data grid standard for gauge
configurations.

\item  implementing new actions such as the Iwasaki gauge action and Overlap
fermions.

\item  building a collection of ready-to-use programs and related
documentation.

\end{itemize}

The development of FermiQCD has greatly benefited from access to other
codes such as UKQCD (from which we borrowed the local random number
generator), MILC (for the staggered fermion algorithms), CANOPY (for many
design features), M.~L\"{u}scher's (for the assembly macros) and
C.~Michael's
(for all-to-all propagators); we thank their authors for making them
available to us. We also wish to thank J.~Flynn, A.~Shams, F.~Mescia, 
L.~Del~Debbio and T.~Rador for their through tests of the inverters 
and the Clover action.

\end{document}